\documentclass[twocolumn,prl,showpacs,preprintnumbers,amsmath,amssymb]{revtex4}
\usepackage{amssymb}
\usepackage{graphics}
\usepackage{epsfig}
\usepackage{graphicx}    % Include figure files
\usepackage{dcolumn}     % Align table columns on decimal point
\usepackage{amsthm}
\usepackage{bm}          % bold math
\usepackage{amsbsy}
\begin{document}

\title{The shape of disorder broadened Landau subbands in graphene}
\author{W. Zhu$^{1}$, Q. W. Shi$^{1 \dagger}$,
X. R. Wang$^{2 \dagger}$, J. Chen$^{3,4}$, J. G. Hou$^{1}$}
\address{$^1$Hefei National Laboratory for Physical Sciences at
Microscale, University of Science and Technology of China, Hefei
 230026, China}
\address{$^2$Department of Physics, The Hong Kong University of
Science and Technology, Clear Water Bay, Kowloon, Hong Kong}
\address{$^3$Electrical and Computer Engineering, University of
Alberta, Alberta, Canada T6G 2V4}
\address{$^4$National Research Council/National Institute of
Nanotechnology, Alberta, Canada T6G 2M9} \email[Electronic
address:]{phsqw@ustc.edu.cn;phxwan@ust.hk}

\date{\today}
\begin{abstract}
Density of states (DOS) of graphene under a high uniform magnetic
field and white-noise random potential is numerically calculated.
The disorder broadened zero-energy Landau band has a Gaussian shape
whose width is proportional to the random potential variance and the
square root of magnetic field. Wegner-type calculation is used to
justify the results.
\end{abstract}
\maketitle

Graphene is a two-dimensional (2D) system that has attracted a lot
of attentions in recent years\cite{a1,a2} because of the distinct
and similar electron properties between graphene and conventional
2D electron gases such as the inversion layers of semiconductor
heterostructures. Different from the usual 2D electrons in solids
that obey Schrodinger equation, the governing equation of electrons
in undoped graphene is the relativistic massless Dirac equation~
\cite{a3} due to the linear dispersion near the vicinity of the
two nonequivalent corners $K$ and $K'$ of the first Brillouin zone.
Similar to the 2D Schrodinger electrons, 2D Dirac electron states
in a perpendicular uniform magnetic field $B$ also form highly
degenerated discrete Landau levels (LLs). Instead of equal spaced
LLs for the Schrodinger electrons, the LLs of 2D Dirac electrons
are unevenly distributed\cite{a10,a13} according to
\begin{eqnarray}
E_n=\pm\sqrt{2e\hbar v_F^2nB}\ \ \ \ \ (n=0,1,2...),
\end{eqnarray}
where $v_F$ is the Fermi velocity, $e$ is electron charge and
$\hbar$ is the Planck constant. The existence of zero-energy Level
for $n=0$ is a direct consequence of charge-hole symmetry. In the
presence of disorders, LLs broaden into Landau subbands, gives raise
to the quantum Hall effects\cite{a15}. The distribution of electron
levels inside the Landau bands is a subject of fundamental interest
because it is important to physics quantities sensitive to the
density of states (DOS). For 2D Schrodinger electrons, the subject
have been studied both theoretically and experimentally for a long
time\cite{a95,a90,a91,a92,a93,a94,c95,c96,c97}. Based on
self-consistent Born approximation (SCBA), the shape of DOS was
argued to be the well-known semicircle form\cite{a95}. Later, Wegner
predicted that the shape of the lowest LL in the strong-field limit
and with white-noise random potential\cite{a90} is Gaussian-like.
The supersymmetry method reaches the similar
conclusions\cite{a92,a94}. However, the shape of disorder broadened
Landau bands for the Dirac electrons is less well studied although
it is important for the issues of diagonal and non-diagonal
conductivity\cite{a50}. In this paper, we show that the shape of
zero-energy Landau band is best described by Gaussian function. The
broadening width $\Gamma$ of the band depends on both the magnetic
field and disorder randomness. In the high magnetic field, it is
proportional to the square-root of the field. At a fixed field, the
broadening width is approximately linear in the square-root of
disorder potential variance. Interestingly enough, these results are
similar to the predictions of Wegner\cite{a90} on Schrodinger
electrons. To demonstrate the importance of the shape of Landau
subband in electron transport in high magnetic field, the
magneto-conductivity is calculated.

%\textit{The Lattice Model and Broadening of LLs}
The $\pi-$electrons of graphene in a perpendicular magnetic field
can be modeled by a tight-binding Hamiltonian on a honeycomb lattice
of two sites per unit cell,
\begin{eqnarray}
H=t(\sum\limits_{i,j}{e^{i\phi_{ij}}|i><j|
+h.c.})+\sum\limits_{i}\varepsilon_i|i><i|
\end{eqnarray}
where $t=-2.7eV$ is the hopping energy between the nearest
neighboring atoms. To mimic randomness on Dirac electrons, the
on-site energy $\varepsilon_i$ is assumed to be random with
Gaussian white-noise distribution of zero mean $<\varepsilon_i>=0$
and finite variance $<\varepsilon_i\varepsilon_j>=\delta_{ij}g^2$.
$g$ measures randomness. The magnetic field $B$ is introduced by
means of Perierls' substitution in hopping parameter\cite{xrw,m1}
as $t\rightarrow te^{i\phi_{ij}}$, where $\phi_{ij}=2\pi e/h
\int_i^j\vec{A}\cdot d\vec{l}$.

In order to obtain numerically an accurate averaged DOS of the above
Hamiltonian, the system size should be large enough so that the
Lanczos recursive method\cite{m2} is used. The averaged DOS can be
evaluated by $\rho(E)=\overline{<\psi| Im\frac{1}{E-H+i\nu}|\psi>}$,
where $\nu$ is an infinitesimal positive number. The bar denotes the
ensemble average. When disorder is present, the system does not have
translational symmetry and all sites are no longer equivalent.
Estimating the average by using the "site-basis" is therefore
inefficient. Inspired by the work of Triozon\cite{m3}, we use the
random phase state\cite{m4}. Thus we only need to consider several
random phase states to get a good estimate of the average DOS. In
this approach, a small artificial cut-off energy ($\epsilon= 1meV$)
is introduced to simulate the infinitesimal imaginary energy
$\nu$\cite{m5,m6}. This artificial parameter will lead to a small
width of LLs in clean graphene.
\begin{figure}
\includegraphics[width=0.5\textwidth]{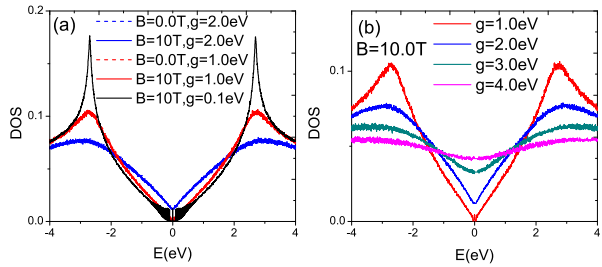}
\caption{(Color online) (a) DOS in a magnetic field of $10T$ (solid
line) and DOS without magnetic field (dashed line) for various $g$.
(b) DOS at $B=10T$ for $g=1, 2, 3, 4$eV. }
\end{figure}

%\textit{Low magnetic-field regime}
Both on-site energy randomness and the magnetic field can affect
DOS. There are two important length scales. Disorder causes electron
scattering, leading to an electron mean-free path $l$. The magnetic
field introduces a magnetic length $l_c=\sqrt{\hbar/eB}$. Electronic
states in a Landau level (LL) can be viewed as cyclotron motions of
an electron around orbits of radius $l_c$ centered at different
location. Scattering can destroy and modify these cyclotron motions,
leading to Landau subband broadening. Thus, one can expect different
DOS at two limits, $l_c\gg l$ and $l_c\ll l$. Fig. 1a is the DOS for
$l_c\gg l$. Indeed, our numerical results show that the DOSs at
$B=10T$ and strong disorder (in the sense of $l_c\gg l$) are almost
the same as DOS at zero field and finite $g$, showing no obvious
magnetic field dependence of the Landau subbands in the limit. Fig.
1b shows DOS of different $g$ at a fixed field $B=10T$ at which
$l_c\gg l$ is still satisfied. The van Hove singularities is less
singular as $g$ increases whereas the DOS near the Dirac point
($E=0$) increases with $g$. These results agree with prior
analytical results\cite{c60}. In the opposite limit (high magnetic
field), the strength of random potential becomes relative weak such
that Dirac electrons can complete their cyclotron motion before
encountering scatters. For $g=0.1$eV, we observed the oscillation of
the DOS near the Dirac point as shown in Fig. 1(a), which reflects
the appearance of LLs in high magnetic field. In what follows, we
focus on the Landau subband broadening in high magnetic field.

%\textit{High magnetic-field regime}
In the limit of $l_c\ll l$, DOS shows a very different behavior.
Fig. 2a is the DOS of graphene without disorder ($g=0$) and
in a uniform magnetic field ($B=10.0T$). $\delta-$like spectrum
distributed according to Eq. (1) is observed. The seemingly
non-zero width is due to the finite size effect and cut-off
energy $1$meV introduced in our numerical calculation.
Furthermore, these finite-size broadened spectrum can be fitted well
by Lorentizian distribution of 2meV width (from the cut-off energy).
This artificial broadening may also be viewed to be a true
broadening accounting the electron-phonon or electron-electron
interactions in reality.

\begin{figure}
\includegraphics[width=0.5\textwidth]{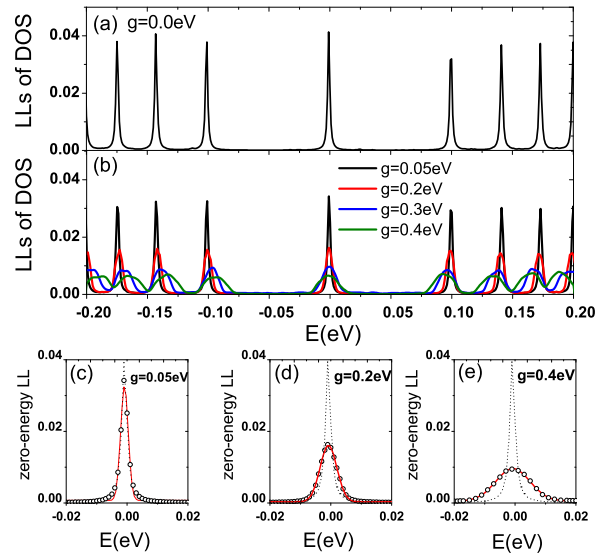}
\caption{(Color online) DOS at $B=10T$. (a) Clean graphene $g=0$;
(b) disordered graphene with $g=0.05$ $0.2,\ 0.3\ 0.4$eV; (c-e)
circles are numerical results of DOS of the zero-energy Landau band
for $g=0.05$eV (c), $g0.2$eV (d), and $0.4$eV (e). The solid (red)
lines are the Gaussian fitting of the form
$\frac{\rho_0}{\Gamma\sqrt{\pi/2}}exp(-\frac{2E^2}{\Gamma^2})$. The
dotted lines are the zero-energy LL of pure graphene. }
\end{figure}

Fig. 2b is the DOS of Landau subbands at $B=10T$ for various $g$.
The LLs degeneracy are lifted in the presence of disorder, LLs are
broadened into Landau subbands. The broadening width ($\Gamma$) of
zero-energy subband increases from $3.0$meV to $11.0$meV when $g$
changes from $0.05$eV to $0.6$eV. When the strength of the
white-noise potential is very weak ($g<0.1$eV), the broadening width
of each Landau band increases a little. As shown in Fig. 2c ($g=
0.05$eV), the shape of the zero-energy Landau band is similar to the
clean case and $\Gamma$ broadened by the disorder is as small as
$3.6meV$. In this case, the dominant factor on the subband shape is
still the artificial cut-off energy. However, the broadening width
$\Gamma$ of zero-energy Landau subband becomes large for $g>0.1$eV
when the disorder dominate the band broadening and the artificial
cut-off energy is irrelevant. Numerical results show that the
broadening shape of zero-energy Landau subband changes also from the
Lorentizian shape to the Gaussian one. Figs. 2d-2e show that DOS can
be fitted well by the Gaussian function, suggesting DOS of
zero-energy Landau subband is Gaussian in the presence of a white
noise potential\cite{e0,e1}. This finding is very similar to the
Gaussian-like shape for the lowest LL for 2D Schrodinger
electrons\cite{a90,a91,a92,a93,a94,c95,c96,c97}.

As shown in Fig. 2b, Landau subbands are distorted and shifted
toward zero energy direction in the presence of white-noise
disorder\cite{c40} except the zero-energy subband because disorder
does not destroy the electron-hole symmetry. This disorder effect
can be seen clearly from the noticeable move of the DOS peak
positions of bands. This shift of DOS peaks can be qualitatively
explained within the Self-consistent non-crossing approximation
(SCNCA)\cite{c50}. According to SCNCA, DOS peak positions should
move toward zero energy in the presence of white-noise disorder due
to the finite real part of the self energies, and the shift
increases with $g$. However, SCNCA fails to account for the shape of
broadened bands.

\begin{figure}
\includegraphics[width=0.5\textwidth]{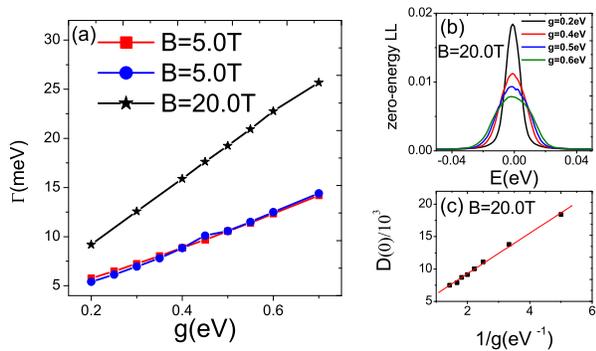}
\caption{(Color online) Disorder-dependence of DOS for fixed $B=5T$
and $B=20T$. (a) $\Gamma$ vs. $g$ for $n=0-$ and $n=1-$bands. It
shows weak (or no) depednence of $\Gamma$ on band index $n$. The
linear relationship suggests $\Gamma\propto g$. (b) DOS of
zero-energy band for various $g$ and at $B=20T$. (c) $D(0)$ vs.
$1/g$ for the zero-energy band at $B=20T$.}
\end{figure}

Fig. 3a shows the $g-$dependence of subband width $\Gamma$ for the
zero-energy subband ($n=0$) for $B=5T$ and $20T$. As expected,
$\Gamma$ of zero-energy Landau subband increases with $g$. The same
dependence is obtained for the first ($n=1$) Landau subband
(squares). The data shows approximately a linear dependence in $g$.
The slop depends on the magnetic field. However, the width does not
depends on the band index $n$ as suggested by the complete overlap
of $\Gamma$ for $n=0$ (circles) and for $n=1$ (squares) at $B=5T$.
Fig. 3b shows the DOS of zero-energy subband at $B=20T$ for various
disorder $g$. The peak $D(0)$ decreases with $g$, and is
proportional to the inverse of $g$ as shown by the good linear fit
in Fig. 3c. This is because the total number of states in each
Landau subband is a constant which is the total number of flux
quanta in the system. Since $\Gamma$ is proportional to $g$, to keep
the area defined by the DOS curve and E-axis to a constant, the peak
must be roughly follow $1/g$ dependence. More interestingly, the
shape of zero-energy subband remains symmetric due to electron-hole
symmetry, and may follow a true Gaussian distribution.

Fig. 4 is the magnetic field dependence of DOS at a fixed $g$. Fig.
4a shows square-root field dependence of $\Gamma$ for both $n=0
-$band and $n=1-$band. The good overlap of $\Gamma$ for the two
subbands at $g=0.2$eV suggests that, similar to the $\Gamma-g$
dependence, $\Gamma-B$ curves are universal (independent on band
indices). Just like DOS-$g$ dependence, Fig. 4b shows how DOS of
zero-energy band change with magnetic field. It is clear that $D(0)$
is proportional to square-root of $B$ as shown in Fig. 4c. This is
because the total number of states in one Landau subband is
proportional to $B$, thus the product of the band peak ($D(0)$) and
bandwidth $\Gamma$ should be proportional to $\sqrt{B}$.

\begin{figure}
\includegraphics[width=0.50\textwidth]{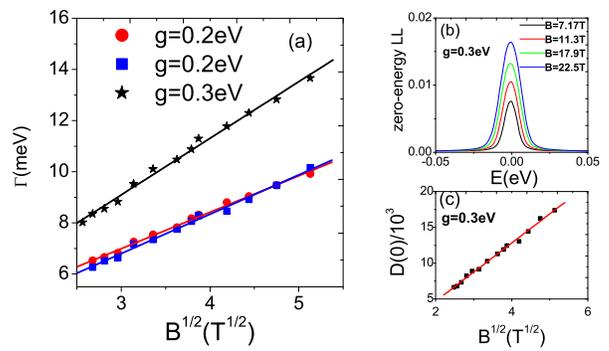}
\caption{(Color online) Field-dependence of DOS for various fixed
$g$. (a) $\Gamma$ vs. $\sqrt{B}$ for $n=0-$ and $n=1-$bands. $g$ is
fixed at $0.2$eV and $0.3$eV. The data shows weak (or no) depednence
of $\Gamma$ on band index $n$. The linear relationship suggests
$\Gamma\propto\sqrt{B}$. (b) DOS of zero-energy band for various $B$
and at a fixed $g=0.3$eV. (c) $D(0)$ vs. $\sqrt{B}$ for the
zero-energy band for $g=0.3$eV.}
\end{figure}

%\textit{Exact density of states for zero-energy LL}
In order to understand our numerical findings quantitatively, we
generalize the approach of Wegner\cite{a90} for Schrodinger
electrons to the Dirac electrons to obtain an analytic expression of
DOS for the zero-energy band since it is only valid for small $E$.
In the high magnetic field regime, the Green's function of
unperturbed system for the zero-energy band is
\begin{eqnarray}
G_0(r,r',E)=<r|\frac{1}{E-H}|r'>=C(r,r')\frac{1}{E-E_0+i\epsilon},
\end{eqnarray}
with
\begin{eqnarray}
C(r,r')=\rho_0exp(-(\xi^*\xi-2\xi'^*\xi+\xi'^*\xi')/(2l_c)^2)
\left(\begin{array}{ccc}
0 & 0  \\
0 & 1  \\
\end{array}\right),\nonumber
\end{eqnarray}
where $\xi=x+iy$ and $\xi'=x'+iy'$ are the complex variables.
$\rho_0=(2\pi l_c^2)^{-1}$ is the degeneracy per area in each LL for
the clean graphene. If the intervalley scattering between $K$ and
$K'$ is ignored, the Green's function of the zero-energy LL has only
one non-zero term, which simplifies the problem a lot. In this case,
the shape of zero-energy LL can be directly mapped to the problem of
conventional 2DEG for the lowest LL\cite{a90}. Therefore, the shape
of the zero-energy band has a Gaussian-like form:
\begin{eqnarray}
\rho(E)=\frac{\rho_0}{\pi}\frac{d}{dE}\tan^{-1}(\frac{2}{\sqrt{\pi}}
\int_0^{\sqrt{E^2/\rho_0g^2}}e^{\eta^2}d\eta).
\end{eqnarray}
This Gaussian-like form is little bit flater than the true Gaussian
distribution. Although our numerical date can also fit this
Gaussian-like form, but they fit the Gaussian function better. This
may be due to the fact that intervalley scattering is neglected in
our derivation of the Gaussian-like form while our numerical
calculation include all effects from both disorders and magnetic
field.

The shape of Landau bands is important to many physical quantities.
As an example, we would like to evaluate the magneto-conductivity of
a band whose DOS is a Gaussian function. In the Drude-Boltzemann
approximation, the Kubo formula yields the diagonal
magneto-conductivity for the Fermi energy at the Dirac
point\cite{d4}:
\begin{eqnarray}
\sigma_{xx} =\frac{4e^2}{h}\frac{te^{-t}}{1-e^{-2t}},
\end{eqnarray}
where $t=\frac{(\hbar\omega_c)^2}{\Gamma^2}$ and $\omega_c=v_F/l_c$
is cyclotron frequency. Although Eq. (4) works well only in the high
magnetic field ($t\geq1$). Noted that $t$ is independent on $B$ as
$\Gamma\propto g\sqrt{B}$, thus magneto-conductivity does not depend
on the magnetic field, and is only disorder dependent. This behavior
reflects that $\sigma_{xx}$ should tend to saturate in high magnetic
field, which is consistent with the experimental observations of
Ref.\cite{a21}.

In conclusion, the shape of disorder broadened zero-energy
Landau band can be described well by a Gaussian function.
For white-noise random potential, band width is proportional to both
the square-root of magnetic field and variance of random potential.
The perturbation theory is used to justify the conclusions.
The importance of the shape of Landau bands is demonstrated by
the magneto-conductivity. The results can shield some lights on
the properties of graphene in magnetic field.

%\textit{Acknowledgement}
This work is partially supported by the National Natural Science
Foundation of China (Grant nos. 10574119). The research is also
supported by National Key Basic Research Program under Grant No.
2006CB922000. X. R. Wang is supported by UGC/RGC grants. Jie Chen
would like to acknowledge the funding support from the National
Institute of Nanotechnology and from the Discovery program of
Natural Sciences and Engineering Research Council of Canada under
Grant No. 245680.


\begin{thebibliography}{99}
\bibitem{m2}Periodic boundary condition are used. Typical supercell size contains $8.0$ millions of carbon atoms.
\bibitem{m4}The random phase states are placed in the middle of the system and extend over a
(possibly large) number of orbitals, Ni:
$|\psi>=\frac{1}{\sqrt{Ni}}\sum\limits_{j}e^{2i\pi\alpha(j)}|j>$,
where $\alpha(j)$ is a random number in the interval $[0, 1]$.
\bibitem{m5}The relationship between $\epsilon$ and the system size ($L$) has
been shown in Ref.\cite{m6}. In the region $\epsilon<10^{-2}$,
$\epsilon$ decreases very slowly when $L$ gets larger. If one wants
to reduce $\epsilon$ to $\epsilon<10^{-3}$, a much large system size
is needed.
\bibitem{e0}We also find other arbitrary uncorrelated short range impurity
potential(such as Anderson disorder) can result in the Gaussian
shape of broadening LLs, which is the same with the results in 2DEG
as well\cite{a15}.
\end{thebibliography}
\end{document}